\documentstyle[12pt,fleqn]{article}
\date{}
\newcommand{\eeq}{\end{eqnarray}}
\newcommand{\beq}{\begin{eqnarray}}

\def\con{{}_{\_\rule{-1pt}{0pt}\_}
\rule{-2pt}{0pt}\raise1.5pt\hbox{$\mid$}\hspace{2pt}}
\newcommand{\tr}{\mbox{\boldmath $\triangle$}}

\newtheorem{TH}{Theorem}
\newtheorem{LM}{Lemma}
\title{$SO(2)$ symmetry of a Maxwell $p$-form theory}
\author{Dariusz Chru\'sci\'nski\footnotemark \\
 Institute of Physics, Nicholas Copernicus University\\
 ul. Grudzi\c{a}dzka 5/7, 87-100 Toru\'n, Poland}

\begin{document}
\def\thefootnote{\relax}\footnotetext{$^*$E-mail:
darch@phys.uni.torun.pl}

\maketitle

\begin{abstract}

We find a {universal} $SO(2)$ symmetry of a $p$-form Maxwell theory
for both odd   
and even $p$. For odd $p$ it corresponds to the duality rotations but for even
$p$ it defines a new set of transformations which is not related to duality
rotations. In both cases a symmetry group defines a subgroup of the $O(2,1)$
group of {\bf R}-linear canonical transformations which has also a
natural representation on the level of quantization condition for $p$-brane
dyons.

\end{abstract}

\section{Introduction}

The old idea of electric-magnetic duality plays in recent years very
prominent role (see e.g. \cite{Olive}). In the present Letter we
investigate this idea in the context of a  Maxwell $p$-form theory in
$D=2p+2$ dimensional Minkowski space-time ${\cal M}^{2p+2}$ (we choose a
signature of the
Minkowski  metric $(-,+,...,+)$). The motivation to study such type
of theories comes e.g. from a string theory where one considers
higher dimensional objects (so called $p$-branes) interacting with a
gauge field.

Let $A$ denote a $p$-form potential
and $F=dA$ be a corresponding $(p+1)$-form field strength. Then the
generalized Maxwell equations are given by:
\beq   \label{M}
d\star F=0\ ,
\eeq
where ``$\star$'' denotes a Hodge operation in ${\cal M}^{2p+2}$. This
operation satisfies
\begin{equation}   \label{*2}
\star\star = (-1)^p\ ,
\end{equation}
which implies a fundamental difference between duality transformations defined
for different parities of $p$. Introducing a complex field
\begin{equation}
X := F + i\star F
\end{equation}
the duality rotations have the following form:
\beq \label{dual-}
X &\rightarrow& e^{-i\alpha}\, X\ ,\ \ \ \ \ \ \ \ \ \ \ \ \ \ \ \ \ \ \ \
\ \ \ \ \ \ \  \
\mbox{for}\ p\ \mbox{odd}\ ,\\
\label{dual+}
X &\rightarrow& \cosh\alpha\, X + i\sinh\alpha\, \overline{X} ,\ \ \ \ \ \
\ \ \ \ \mbox{for}\ p\
\mbox{even}\ ,
\eeq
with $\overline{X}$ denoting a complex conjugation of $X$. Note, that there is
an essential  difference between analytical properties of $SO(2)$
rotations (\ref{dual-}) and hyperbolic
$SO(1,1)$ rotations (\ref{dual+}). Namely, the former are {\bf C}-linear
whereas the latter are only {\bf R}-linear. This mathematical difference
has even stronger physical consequences.
It is well known that  Maxwell  $p$-form electrodynamics is duality
invariant only for odd $p$   \cite{Gibbons}, \cite{Deser1}, i.e. only
(\ref{dual-}) defines a symmetry of a theory.

In the present Letter we show that there is a universal $SO(2)$ symmetry
of a $p$-form Maxwell theory which is
valid for any $p$. When $p$ is odd this symmetry is equivalent to
(\ref{dual-}) but when $p$ is even it does not correspond to the duality
rotation (\ref{dual+}).

In order to find this new symmetry we look more
closely into the hamiltonian structure of a $p$-form theory. Decomposing ${\cal
M}^{2p+2}$ into a time-line and a $(2p+1)$ dimensional space-like hyperplane
$\Sigma$  the Maxwell equations (\ref{M}) supplemented by the Bianchi identity
$dF=0$ have the following form:
\begin{equation}   \label{Mp}
\dot{{\cal E}} = dB\ ,\ \ \ \ \ \dot{{\cal B}}= (-1)^p\, dE\ ,
\end{equation}
together with the Gauss constraints
\begin{equation}      \label{Gauss}
d{{\cal E}} =0\ ,\ \ \ \ \ d{\cal B}=0\ ,
\end{equation}
where $E$ and $B$ are electric and magnetic
$p$-forms on $\Sigma$ defined in the standard
way {\em via} $F$ (see e.g. \cite{Gibbons}).
The
dual $(p+1)$-forms ${\cal E}=\star E$ and ${\cal B}=\star B$ are defined {\em
via} the Hodge star $\star$ on $\Sigma$ induced form ${\cal M}^{2p+2}$ (we
shall use the same symbol for both operations). Note the presence of a
$p$-dependent sign in (\ref{Mp}) which plays a crucial role in what follows.

Now, our strategy looks as follows: in the next section we define new
variables $Q$'s and $\Pi$'s which are more convenient for our purpose. We call
them reduced variables because they already solve the Gauss constraints
(\ref{Gauss}). In Section~\ref{Can} we reformulate the canonical structure of
a $p$-form theory in terms of the reduced variables. It allows us to make an
observation that theories with different parities of $p$ are related by a
simple transformation of variables. It turns out that the canonical
structure possesses a natural $O(2,1)$ invariance group. For odd $p$ there
is a
$SO(2)$ subgroup of $O(2,1)$ corresponding to duality rotations
(\ref{dual-}). However, for even $p$
no such a subgroup corresponding to
$SO(1,1)$ rotations (\ref{dual+}) exists. Nevertheless, also in this case,
there is a $SO(2)$ subgroup defining a symmetry of a Maxwell theory.
This subgroup is a true
counterpart of (\ref{dual-}) for a theory with $p$ even.
 Finally,  we show that there is a natural realization of
the above $O(2,1)$ symmetry on a level  of the generalized Dirac
quantization condition for $p$-brane dyons \cite{dyon}.

\section{Reduced variables}

Let us choose spherical coordinates on
$\Sigma$ centered at some arbitrary point and let $S^{2p}(r)$ denote a $2p$
dimensional sphere of radius $r$. Using the canonical embedding
\beq
\phi_r\ :\ S^{2p}(r) \rightarrow \Sigma
\eeq
let us define the following $(p-1)$-forms on each $S^{2p}(r)$:
\beq
Q_1 &:=& \phi_r^* \, ( \partial_r \con E)\ , \\
Q_2 &:=& \phi_r^* \, ( \partial_r \con  B)\ ,
\eeq
and
\beq
\Pi_1 &:=& r \tr_{p-1}^{-1} \, d\, \star\, \phi^*_r\, B\ , \\
\Pi_2 &:=& - r \tr_{p-1}^{-1} \, d\, \star\, \phi^*_r\, E ,
\eeq
where $\tr_{p-1}$ denotes a Laplacian on $(p-1)$-forms on
$S^{2p}(1)$. In the above formulae ``$d$'' and ``$\star$'' are operations on
$S^{2p}(r)$ induced from $\Sigma$ {\em via} $\phi_r$. Note, that $\tr_{p-1}$
is invertible (cf. \cite{ROMP}), therefore, $\Pi$'s are well defined.

The crucial property of the above defined variables consists in the following
\begin{TH} 
The quantities $(Q_\alpha,\Pi_\alpha);\ \alpha=1,2$
contain the entire gauge-invariant information about the fields $E$
and $B$. Moreover, they already solve Gauss constraints (\ref{Gauss}).
\end{TH}
For proof see \cite{ROMP}. Note, that $\Pi$'s are highly nonlocal functions
of $E$ and $B$. However, it was observed long ago \cite{DESER} that in
ordinary ($p=1$) electrodynamics duality rotations (\ref{dual-}) are generated
by a nonlocal operator.  Therefore, as we show, $Q$'s and $\Pi$'s are well
suited to study the duality invariance of a $p$-form theory. More detailed
analysis of these variables may be found in \cite{ROMP}.

\section{Canonical structure}
\label{Can}

Having any two $k$-forms $\alpha$ and $\beta$ on $S^{2p}(r)$ let us define
a standard scalar product:
\beq
(\alpha,\beta)_r := \int_{S^{2p}(r)} \alpha \wedge \star \beta\ .\nonumber
\eeq
Moreover, let
\beq
(\alpha,\beta) := \int_{0}^{\infty} dr\ (\alpha,\beta)_r\ .\nonumber
\eeq
With this notation one has

\begin{TH}
The phase space of a $p$-form theory is endowed with the
canonical symplectic structure $\Omega_p$ given by:
\beq
\Omega_p = (\delta\Pi_1,\wedge \delta Q_1) + (-1)^{p+1}(\delta
\Pi_2,\wedge \delta Q_2)\ .
\eeq
\end{TH}
For proof see \cite{ROMP}. Now, let us define complex forms:
\beq
Q &:=& Q_1 + iQ_2\ ,\\
\Pi &:=& \Pi_1 + i\Pi_2\ .
\eeq
Denoting by $\Omega_-$ ($\Omega_+$) a symplectic form $\Omega_p$ for odd
(even) $p$  one has:
\beq  \label{Omega-}
\Omega_- &=& \mbox{Re}\ (\delta\overline{\Pi},\wedge\delta Q)\ ,\\
\label{Omega+}
\Omega_+ &=& \mbox{Re}\ (\delta\Pi,\wedge\delta Q)\ ,
\eeq
where  ``Re''
stands for a real part. Finally, Maxwell equations rewritten  in terms of
reduced variables read (cf. \cite{ROMP}):
\beq
\dot{Q} &=& - \tr_{p-1}\ \Pi\ ,\nonumber\\
\dot{\Pi} &=& - \tr_{p-1}^{-1}\left[ r^{-1}
\partial^2_r(rQ) + r^{-2}
\tr_{p-1}\,Q
\right]\ ,   \label{Max-}
\eeq
for odd $p$, and
\beq
\dot{Q} &=& - \tr_{p-1}\ \overline{\Pi}\ ,\nonumber\\
\dot{{\overline{\Pi}}} &=& - \tr_{p-1}^{-1}\left[ r^{-1}
\partial^2_r(rQ) + r^{-2}
\tr_{p-1}\,Q
\right]\ ,   \label{Max+}
\eeq
for even $p$. One  easily shows that
 (\ref{Max-}) and (\ref{Max+}) define
hamiltonian equations  with respect to $\Omega_-$ and $\Omega_+$
respectively generated by the following Hamiltonian:
\beq              \label{H-Maxwell}
H_p &=& \frac{1}{2(p-1)!}  \left[ \left(r^{-1}Q,r^{-1}\overline{Q}\right)
 - \left(r^{-1}\partial_r(r
Q),\tr_{p-1}^{-1}r^{-1}\partial_r(r\overline{Q})\right)
 - \left(\Pi,\tr_{p-1}\overline{{\Pi}}\right) \right]\ .
\eeq
Moreover, one may easily prove
\begin{LM}
Numerically $H_p$ equals to the standard Maxwell Hamiltonian obtained {\em
via} the symmetric energy-momentum tensor
\beq
H_p = \frac{1}{2p!}\int_{\Sigma} (E \wedge {\cal E} +  B \wedge {\cal B} )\ .
\nonumber
\eeq
\end{LM}
Note, that the difference between theories with different parities of $p$
is now very transparent. Namely, they are related by a
simple replacement $\Pi \rightarrow \overline{\Pi}$. This way one obtains
(\ref{Omega+}) from (\ref{Omega-}) and (\ref{Max+}) from
(\ref{Max-}). Note, that the Hamiltonian (\ref{H-Maxwell}) is invariant under
$\Pi \rightarrow \overline{\Pi}$.

\section{Canonical transformations and duality rotations}

Now, let us look for the canonical transformation with respect to
$\Omega_-$ and $\Omega_+$. In the class of {\bf R}-linear transformations we
have
\begin{LM}
The following generators:
\beq
G_1 = {\em Im}\ (Q,\overline{\Pi})\ ,\ \
G_2 = {\em Im}\ (Q,{\Pi})\ ,\ \
G_3 = {\em Re}\ (Q,\overline{\Pi})\ ,\ \
G_4 = {\em Re}\ (Q,{\Pi})\ , \nonumber
\eeq
generates both with respect to $\Omega_-$ and $\Omega_+$ the $O(2,1)$
group of {\bf R}-linear canonical transformations.
\end{LM}
Note, that the set $(G_1,G_2,G_3,G_4)$ is closed with respect to
$\Pi\rightarrow \overline{{\Pi}}$. Moreover, there is $O(2)$ subgroup of 
{\bf C}-linear transformations generated by $(G_1,G_3)$ and $(G_2,G_4)$
for odd and even $p$ respectively (actually, $G_3$ and $G_4$ generate the
corresponding centers of $O(2,1)$).

Now, it is easy to prove the following
\begin{TH}
The Maxwell Hamiltonian (\ref{H-Maxwell}) is invariant under the action of
$G_1$ and $G_2$ for odd and even $p$ respectively.
\end{TH}

Let us observe that the duality rotations (\ref{dual-}) and (\ref{dual+})
may be expressed as follows:
\beq
Q &\rightarrow& e^{-i\alpha} \, Q\ ,\nonumber\\
\Pi &\rightarrow& e^{-i\alpha} \, \Pi\ ,  \label{D-}
\eeq
for $p$ odd, and
\beq
Q &\rightarrow& \cosh\alpha\ Q - i\sinh\alpha\ \overline{Q}\ ,\nonumber\\
\Pi &\rightarrow& \cosh\alpha\ \Pi + i\sinh\alpha\ \overline{{\Pi}}\ ,
\label{D+} 
\eeq
for even $p$.
One immediately sees that (\ref{D-}) are generated by $G_1$ but none of
$G_k$ does correspond to (\ref{D+}).
This fact that
the hyperbolic $SO(1,1)$ rotations are not even implementable as
canonical transformations was observed in a slightly different context in
\cite{Deser1}.

Therefore, the  true counterpart of (\ref{D-}) for  $p$ even is not
(\ref{D+}) (it is obvious because they are not related {\em via}
$\Pi\rightarrow \overline{\Pi}$) but
\beq
Q &\rightarrow& e^{-i\alpha} \, Q\ ,\nonumber\\
{\Pi} &\rightarrow& e^{i\alpha} \, {\Pi}\ ,  \label{new}
\eeq
which is generated by $G_2$ {\em via} $\Omega_+$.

Finally, let us note that
quantum mechanics applied to a $p$-form theory implies the following
quantization condition for  $p$-brane dyons \cite{dyon}:
\beq  \label{q}
e_1g_2 + (-1)^p e_2g_1 = nh\ ,
\eeq
with an integer $n$ ($h$ denotes the Planck constant). For odd $p$ the
above condition is a generalization 
of the famous Dirac condition \cite{Dirac} but for even $p$ it was
observed only recently \cite{dyon}. Again, a parity of
$p$ plays a crucial role in (\ref{q}).
Introducing a complex charge
\beq
q := e + ig  \nonumber
\eeq
the formula (\ref{q}) may be rewritten as follows:
\beq   \label{q-}
\mbox{Im}\ (\overline{q}_1 q_2) &=& nh\ ,\ \ \ \ \ \ \ \ p \ \mbox{odd}\ ,\\
\label{q+}
\mbox{Im}\ (q_1{q}_2) &=& nh\ ,\ \ \ \ \ \ \ \ p \ \mbox{even}\ .
\eeq
Let us observe that there is a direct correspondence  between formulae
(\ref{Omega-}) and (\ref{q-}) and formulae  (\ref{Omega+}) and (\ref{q+}).
Therefore, making the following replacements: $\Pi\rightarrow q_1$ and
$Q\rightarrow q_2$ we obtain the natural action of $O(2,1)$  on the level
of charges.

\end{document}